\documentclass[10pt,a4paper]{article}

\usepackage[T1]{fontenc}
\usepackage[utf8]{inputenc}
\usepackage{lmodern}
\usepackage{microtype}
\usepackage[margin=2.4cm]{geometry}
\usepackage{amsmath,amssymb,amsthm}
\usepackage{graphicx}
\usepackage{booktabs}
\usepackage{setspace}
\usepackage[labelfont=bf,font=footnotesize]{caption}
\usepackage[numbers,sort&compress]{natbib}
\usepackage{xcolor}
\usepackage{xurl}
\usepackage[colorlinks=true,linkcolor=black,citecolor=blue!50!black,
            urlcolor=blue!50!black]{hyperref}

\usepackage{mathtools}
\usepackage{bm}
\usepackage{units}
\graphicspath{{figures/}}

\newcommand{\PaperTitle}{Vibro-acoustic optimisation of sandwich panels using the wave/finite element method}
\newcommand{\PaperAuthors}{C. Droz$^{1}$, Z. Zergoune$^{1,2}$, R. Boukadia$^{1}$, O. Bareille$^{1}$ and M.N. Ichchou$^{1}$}
\newcommand{\PaperAffil}{%
  $^{1}$\'Ecole Centrale de Lyon, 36 Avenue Guy de Collongue, 69134 \'Ecully Cedex, France\\
  $^{2}$Facult\'e des Sciences et Techniques de F\`es, Laboratoire de G\'enie M\'ecanique, Route d'Immouzzer, 2202 F\`es, Morocco}
\newcommand{\PaperDate}{Accepted manuscript, originally 2016}
\newcommand{\PaperKeywords}{Honeycomb, Finite element, Sandwich, Wave, Acoustic, Design}

\newcommand{\PaperDOI}{10.1016/j.compstruct.2016.01.025}
\newcommand{\HalID}{hal-03372973}
\newcommand{\ArxivID}{}          

\newcommand{\CiteAs}{%
  C. Droz, Z. Zergoune, R. Boukadia, O. Bareille, M.N. Ichchou, \emph{\PaperTitle}, Composite Structures 156 (2016) 108-114, doi:10.1016/j.compstruct.2016.01.025}

\makeatletter
\newcommand{\HasDOI}{%
  \ifx\PaperDOI\@empty\expandafter\@secondoftwo\else\expandafter\@firstoftwo\fi}
\newcommand{\HasArxiv}{%
  \ifx\ArxivID\@empty\expandafter\@secondoftwo\else\expandafter\@firstoftwo\fi}
\makeatother

\providecommand{\doi}[1]{\href{https://doi.org/#1}{doi:#1}} 

\begin{document}

\begingroup
\renewcommand{\thefootnote}{\fnsymbol{footnote}}
\begin{center}
{\LARGE\bfseries \PaperTitle\par}
\vspace{0.7em}
{\normalsize \PaperAuthors\footnotemark[1]\footnotemark[2]\par}
\end{center}
\footnotetext[1]{\mbox{}\\[-0.7em]\raggedright\PaperAffil}
\footnotetext[2]{\PaperDate.
  Author's version with AI-assisted typesetting and layout. Not the publisher version of record.
  \HasDOI{Definitive publication: \doi{\PaperDOI}.}{No journal DOI yet.}
  HAL: \href{https://hal.science/\HalID}{\HalID}.
  \HasArxiv{arXiv: \href{https://arxiv.org/abs/\ArxivID}{\ArxivID}.}{}}
\endgroup
\setcounter{footnote}{0}

\vspace{0.85em}
{\small
\noindent\textit{How to cite this version.}\\[0.35em]
\raggedright \CiteAs
\par}
\vspace{1.05em}

\begin{abstract}
\noindent
This paper investigates the use of a wave-based method in the framework of structural optimisation of composite panels involving advanced components. The wave/finite element method (WFEM) is used to evaluate the influence of a core's geometry on the transition frequency of a sandwich panel involving composite skins. This transition occurs in a sandwich panel when the transverse shear stiffness has a significant influence on the flexural motion, compared to the bending stiffness. It follows that the modal density and the acoustic radiation will considerably increase above this frequency. The periodic waveguide is modelled at the mesoscopic scale using a 3D finite element model of the unit-cell. Therefore this method does not require a homogenisation of the core based on Gibson and Ashby formulations to provide the wave dispersion characteristics. Although the cellular cores compared in this study share the same mass-to-stiffness ratio, a significant alteration of the transition frequency and modal density can be observed compared to honeycomb cores. A periodic octagonal core is designed, providing up to 70\% increase of the transition frequency and a significant reduction of the modal density.
\end{abstract}

\noindent\textit{Keywords:} \PaperKeywords


\section{Introduction}

Sandwich panels are extensively used in automotive, naval and aerospace industry. These structures have high stiffness-to-weight ratios and are usually made of a moderately thick, lightweight core surrounded by glass or carbon-fibre composite skins. The core can be made of polymer foam or more complex structures involving aluminium or resin honeycomb constructions. Although the material and geometrical properties of the panel are usually designed to provide specific stiffness and density characteristics, structural optimisation also concerns acoustic radiation efficiency in order to improve the acoustic comfort. Therefore, there is an increasing need for reliable optimisation tools for design engineers, providing fast vibro-acoustic evaluation of large-scale structures involving composite components. 

In this context, the knowledge of accurate wave dispersion characteristics in two laminated orthotropic skins connected by 3D cellular honeycomb core is a key information for the prediction of the acoustic transmission parameters. Numerical methods for analysing the vibrational behaviour of complex composite or periodic panels in a broadband frequency range were extensively investigated in the last decade. Honeycomb sandwich panels were usually modelled using classical laminate plate theory (CLPT) or Reissner--Mindlin (FSDT) theories involving three homogeneous layers. An extensive number of analytical formulas for analysing wave dispersion characteristics in two-dimensional waveguides can be found in the literature (see Reddy \cite{Reddy1997}), starting with the asymptotic model for analysing wave dispersion in thick symmetric sandwich panels developed by Kurtze and Watters \cite{Kurtze1959}, then later in the classic work of Wilkinson \cite{Wilkinson1968} and Erickson \cite{Erickson1969}. The prediction of wave propagation characteristics of structurally advanced structures using analytical models has been a subject of intense research over the past years. Renji et al. \cite{Renji1996} introduced a new analytical formulation providing a core transverse shear stiffness and bending matrices in both directions for orthotropic honeycomb sandwich panels. Ghinet and Atalla \cite{Ghinet2011} also used an analytical model based on Discrete Laminate Theory (DLT) to compute the dispersion characteristics of layered structures and obtain their acoustic parameters. 

Recently, Guillaumie \cite{Guillaumie2015} proposed an analytical solution for the eigenmodes and modal densities of symmetric sandwich panels involving composite skins and honeycomb core. This model was found to be effective in the low- and medium-frequency range for moderately thick sandwich structures and was confirmed by experiments. Florence and Renji \cite{Florence2015} also derived explicit solutions for the modal density of composite cylinders.

However, these analytical solutions require a homogenisation of the hexagonal cells of the core based on Gibson and Ashby formulations \cite{Gibson1997}. These analytical approaches therefore suffer some drawbacks in the medium-frequency range, since they cannot handle arbitrarily shaped 3D cells for the core. Besides, these models are often inaccurate when compared to experimental characterizations in higher frequency range, especially when dealing with complex industrial products involving thick, multi-layered constructions made of anisotropic glass or carbon fibres skins. 

Therefore, numerical methods were recently developed to perform wave analysis in complex periodic waveguides defined using classical finite element packages (FEM). Among others, the wave finite element method (WFEM) \cite{MaceJASA2005,MencikEJM2005} has been successfully applied to predict the radiation efficiency of numerous thick layered structures. Nevertheless, these models involve homogeneous orthotropic layers for the honeycomb construction, and may become inaccurate when the wavelengths are reaching the periodic cells dimensions. Although the design optimisation of honeycomb cells has received a lot of attention recently \cite{Alkhader2015,Catapano2014}, the dispersion characteristics of sandwich panels involving realistic composite skins and arbitrarily shaped cores were not investigated to the author's knowledge.

This paper focuses on the influence of the meso-scale parameters of the core's construction on the vibro-acoustic response of the sandwich panel. A special interest is given to the modal density, a typical vibro-acoustic parameter and the transition frequency, associated with the passage from a behaviour where the flexural wave is ruled by the skins stiffness to a behaviour where the core's transverse shear parameters govern the flexural motion. First, finite element models are developed for the periodic cells. The WFEM, which combines FEM and the Periodic Structures Theory (PST) is applied to evaluate the wave dispersion characteristics. Since numerous degrees of freedom are usually involved to ensure accuracy in a broadband frequency range, reduced order modelling (ROM) was recently investigated for periodic \cite{Ruzzene2016,Zhou2015CS} and large-scale \cite{DrozCS2014} waveguides. A hybrid wave-mode ROM is therefore applied to provide suitable computational efficiency. When dealing with hexagonal cores, the equivalent orthotropic material can be described using Gibson and Ashby formulations, and the analytical expression of the wavenumber can be derived for the symmetric sandwich. The possibility to use a regression on the wavenumber function is therefore examined for various core's configurations, in order to derive an analytical expression of the transition frequency from the equivalent bending and shear parameters. Finally, the effects of several geometrical parameters of the core on the modal density and the transition frequency are investigated.

\section{Wave propagation in 2D periodic structures \label{Sec2}}

	\subsection{The wave finite element method}
	
\begin{figure}[ht]
	\begin{center}
	\includegraphics[width=0.35\linewidth]{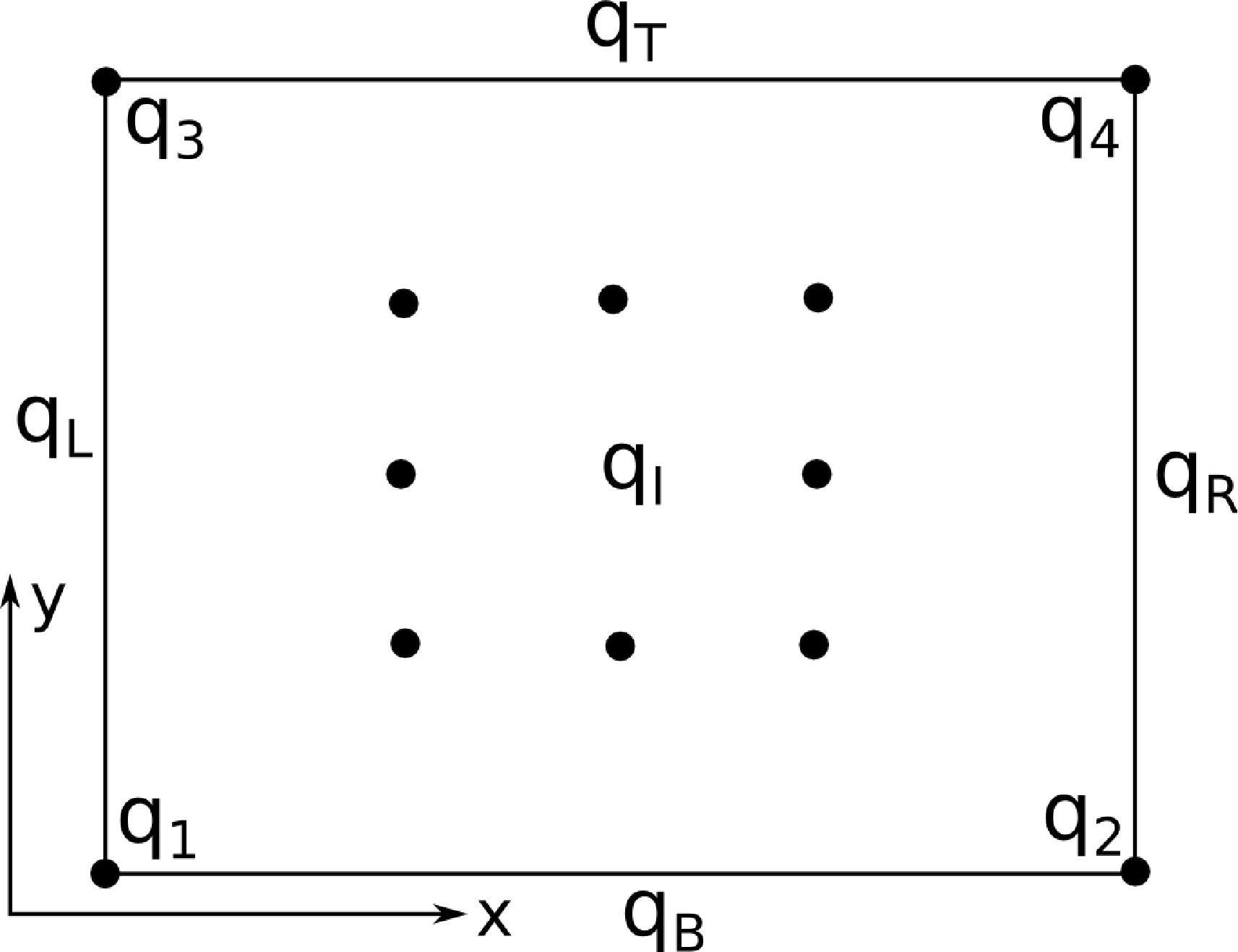}
	\end{center}
	\caption{Top-view of the modelled periodic element with its edge (B, T, R, L), corner (1, 2, 3, 4) and inner (I) degrees of freedom.}
	   \label{periodic_cell}
\end{figure}

Numerous formulations of the WFEM were given in the literature. The method is briefly reviewed hereby. Consider a periodic cell of a 2D waveguide of dimensions $d_x$ and $d_y$ in x and y directions (see Fig. \ref{periodic_cell}), modelled using a FE software. The mass $\mathbb{M}$, stiffness $\mathbb{K}$ and damping $\mathbb{C}$ matrices can be extracted and the dynamic stiffness matrix is written at the circular frequency $\omega$: $\mathbb{D} = \mathbb{K} + j \omega \mathbb{C} - \omega^2 \mathbb{M}$. The degrees of freedom $\mathbf{q}$ (DOF) on the edges, corner and centre of the cell can be reordered as follows:
\begin{equation}
	\mathbf{q} = \left[\mathbf{q}_I^T \quad \mathbf{q}_B^T \quad \mathbf{q}_T^T \quad \mathbf{q}_L^T \quad \mathbf{q}_R^T \quad \mathbf{q}_1^T \quad \mathbf{q}_2^T \quad \mathbf{q}_3^T \quad \mathbf{q}_4^T \right]^T.
	\end{equation}
	
	It should be noted that apart from the mesh compatibility between the opposite edges of the unit-cell, the discretization is arbitrary inside the periodic element. Then using the periodicity relations and Bloch's theorem \cite{Bloch1929}, a reduced state vector can be defined using the propagation constants $\lambda_x$ and $\lambda_y$ in x- and y-directions:
	\begin{equation}
	\left( \begin{array}{c}
	\mathbf{q}_I\\
	\mathbf{q}_B\\
	\mathbf{q}_T\\
	\mathbf{q}_L\\
	\mathbf{q}_R\\
	\mathbf{q}_1\\
	\mathbf{q}_2\\
	\mathbf{q}_3\\
	\mathbf{q}_4\\
	\end{array} \right) =
	\left(\begin{array}{cccc}
	\mathbf{I} & 0 & 0 & 0\\
	0 & \mathbf{I} & 0 & 0\\
	0 & \mathbf{I}\lambda_y & 0 & 0\\
	0 & 0 & \mathbf{I} & 0\\
	0 & 0 & \mathbf{I}\lambda_x & 0\\
	0 & 0 & 0 & \mathbf{I}\\
	0 & 0 & 0 & \mathbf{I}\lambda_x\\
	0 & 0 & 0 & \mathbf{I}\lambda_y\\
	0 & 0 & 0 & \mathbf{I}\lambda_x\lambda_y
	\end{array} \right)
	\left( \begin{array}{c}
	\mathbf{q}_I\\
	\mathbf{q}_B\\
	\mathbf{q}_L\\
	\mathbf{q}_1
	\end{array} \right) = \mathbf{R} \mathbf{q}'.
	\end{equation}
	
	Then, denoting $k_x$, $k_y$ the wavenumbers in $x$ and $y$ directions, the propagation constants $\lambda_x = e^{-jk_xd_x}$ and $\lambda_y = e^{-jk_yd_y}$ are the solutions of the problem:
	\begin{equation}
	\mathbf{R}^T(\lambda_x,\lambda_y) \mathbb{D} \mathbf{R}(\lambda_x,\lambda_y) \mathbf{q}' = \mathbf{0}.
	\label{eq_quadratic}
	\end{equation}
	
	These solutions describe the propagating and evanescent waves in the periodic waveguide. Each solution ($k_x, k_y, \omega$) is associated with an eigenvector $\mathbf{\Psi}$ describing the deformed shape of a given wave. The propagating and evanescent waves can be discriminated from the imaginary parts, since $(\Im(k_x) , \Im(k_y))$ describes wave attenuation in $x$- and $y$-direction, respectively. In dissipative waveguides, all the waves are decaying and a customised selection procedure, usually based on the value of $\dfrac{|\Im(k)|}{\Re(k)}$ can be employed.

	\subsection{Resolution and post-processing}
	
	 It should be noted that Eq.(\ref{eq_quadratic}) is a transcendental eigenvalue problem whose resolution is not straightforward when internal DOFs are involved. In order to turn this problem into a classical eigenvalue problem, two parameters are usually given (the circular frequency $\omega$ and the wavenumber in one direction: $k_x$ or $k_y$). The directional wavenumber $k(\theta, \omega)$ can be obtained from the discrete $(k_x(\omega), k_y(\omega))$ solutions. An accurate description of the 2D dispersion curves (also referred to as \textit{k-space}) hence requires an important number of resolutions of Eq.(\ref{eq_quadratic}). 
	
 In order to enable fast design optimisation, through broadband wave analyses of finely meshed periodic honeycomb constructions, a reduced formulation of the WFEM (see \cite{Droz2015}) is used in the numerical applications. It employs a projection of the state vector $\mathbf{q}'$ on a solution subset involving a reduced number of wave solutions and component modes of the periodic cell. 
	
The flexural waves can then be discriminated using a wave-matching algorithm. At each frequency step, the wave shape is compared to the whole set of propagating waves using a MAC criterion. It expresses the correlation between two vectors $\mathbf{\Phi_i}$ and $\mathbf{\Phi_j}$ by:
	\begin{equation}
	\mathrm{MAC} =  \dfrac{(\Phi_i^t \overline{\Phi_j}) (\Phi_j^t \overline{\Phi_i})}{(\Phi_i^t \overline{\Phi_i}) (\Phi_j^t \overline{\Phi_j})}.
	\end{equation}
	
	Therefore, the wavetypes can be sorted and the flexural wave is identified by the transverse vertical motion of the unit-cell. Although this procedure is not always necessary in lower frequencies it becomes essential for the identification of the flexural wave in higher frequencies, where numerous high-order (or structural) waves can be subjected to veering effects and conversions.
	
	Generally, refined angular sampling for $k(\theta, \omega)$ is necessary for orthotropic or anisotropic structures, which may exhibit high directivity. Therefore, the modal density can be written using classical Statistical Energy Analysis (SEA) expression \cite{Lesueur1988}:
\begin{equation}
n(\omega) = \int_0^{\pi} n(\theta,\omega) d\theta,
\end{equation}
where $n(\theta,\omega)$ denotes the angular distribution of the modal density:
\begin{equation}
n(\theta,\omega) = \dfrac{A}{2\pi^2} \dfrac{k(\theta,\omega)}{|c_g(\theta,\omega)|}.
\end{equation}

$A$ is the surface area and $c_g(\theta,\omega) = \dfrac{\partial \omega}{\partial k}$ is the group velocity in the direction $\theta$. An equivalent wavenumber $\left\langle k_f(\omega)\right\rangle$ is defined so that the modal density can be written:
\begin{equation}
n(\omega) = \dfrac{A}{2\pi} \left\langle k_f(\omega)\right\rangle \dfrac{\partial \left\langle k_f(\omega)\right\rangle}{\partial \omega}.
\end{equation}
	
This expression enables the comparison of scalar solutions of the dispersion relation while ensuring the accurate computation of the modal density in the 2D structure, since the directivity is taken into account. The averaged wavenumber can be expressed in terms of the angular wavenumber:
\begin{equation}
\left\langle k_f(\omega)\right\rangle = \sqrt{\dfrac{1}{2\pi}\int_0^{2\pi} k_f^2(\theta, \omega) d\theta}.
\label{eq_moyennek}
\end{equation}

The modal density can be also determined by the area integration of the wavenumber space (see Langley \cite{Langley1994}). In case of an elliptic orthotropy along x- and y-directions, the equivalent wavenumber can be defined so that the k-space areas remain the same: $\pi k_x k_y = \pi \left\langle k_f \right\rangle^2$. It yields $\left\langle k_f \right\rangle = \sqrt{k_x k_y}$. 

For the sake of clarity, the equivalent flexural wavenumber $\langle k_f(\omega) \rangle$ and group velocity $\frac{\partial \omega}{\partial \langle k_f(\omega)\rangle}$ are simply denoted $k$ and $c_g$ in the following.

	\subsection{Evaluation of the transition frequency}
	
	In sandwich structures, one can discriminate between three frequency domains associated with different kinetic energy distributions. In low frequency, the structure has a classical isotropic plate behaviour. The asymptotic wavenumber modulus $k(f \rightarrow 0)$ and modal density $n(f \rightarrow 0)$ are written:
	\begin{equation}
	k(f \rightarrow 0) = \left( \dfrac{m_s}{D} \right)^{1/4} \sqrt{\omega} \quad ; \quad n(f \rightarrow 0) = \dfrac{A}{2}  \sqrt{\dfrac{m_s}{D}}
	\label{eq_transitionK}
	\end{equation}
	where $A, D, m_s$ denote the area, equivalent bending rigidity and mass per unit area of the plate, respectively.
	
	In a medium frequency (usually acoustic) range, the transverse vertical shear rigidity $S$, increases and influences the dynamical behaviour. The asymptotic solutions for the orthotropic plate are:
	\begin{equation}
	k(f)_\textrm{M.F.} \simeq \omega \sqrt{\dfrac{m_s}{S}}  \quad ; \quad n(f)_\textrm{M.F.} \simeq A  \left( \dfrac{m_s}{S} \right)  \omega
	\label{eq_transitionN}
	\end{equation}

	In terms of waves, higher frequencies are associated with fully localised propagation, meaning that flexural waves can propagate independently in both skins. The transition frequency considered here refers specifically to the passage from the isotropic plate behaviour to a motion governed by the out-of-plane shear rigidity. In Guillaumie \cite{Guillaumie2015}, this frequency is calculated as the intersection between the asymptotic wavenumbers: $k(f \rightarrow 0) $ and $k(f)_\textrm{M.F.} $. The influence of a core's shear rigidity on the modal density and transition frequency of a sandwich plate was well analysed by Renji et al. \cite{Renji1996}. Above the transition frequency the modal density increases linearly with frequency and produces higher levels of acoustic radiation. In the following, the possibility to shift the transition frequency by using different cores geometries is investigated.

\section{Wave dispersion characteristics in arbitrary shaped sandwich panel}

	\subsection{Material properties and geometry of the core}
	
	The periodic panel considered here is a symmetric sandwich plate involving composite faces made of carbon reinforced epoxy resin. The core is made of Nomex paper (aramid fibre and phenolic resin) of thickness 76.2\,$\mu$m assumed isotropic. The skins are made with two plies of transverse isotropic, 0.25\,mm-thick, unidirectional composite (UD) with the stacking sequence [0$^o$, 90$^o$]. The thickness of the core is $h_T$ = 25\,mm and the material properties are detailed in Table \ref{table_mat}.
\begin{table}[!ht]
	\begin{center} 
	\begin{tabular}{|lcccc|}
	\hline
	Material			    &  Density (kg.m$^{-3}$)	&	Young modulus (Pa)	& Shear modulus (Pa)	&	Poisson ratio  \\
	\hline 
	Nomex core			&  $\rho_c$ = 1240	&	E$_c$ = 5.5 $\times$ 10$^9$			& G$_c$ = 2.7 $\times$ 10$^9$			&	$\nu_c$ = 0.33  \\
	\hline
				    			&  							&	E$_1$ = 133.6 $\times$ 10$^9$	& 	G$_{23}$ = 2.6 $\times$ 10$^9$	&	$\nu_{23}$ = 0.49  \\
	UD laminate			    &  $\rho_s$ = 1550	&	E$_2$ = 7.7 $\times$ 10$^9$		& 	G$_{13}$ = 3.1 $\times$ 10$^9$	&	$\nu_{13}$ = 0.29  \\
				    			&  							&	E$_3$ = 7.7 $\times$ 10$^9$		& 	G$_{12}$ = 3.1 $\times$ 10$^9$	&	$\nu_{12}$ = 0.29 \\
	\hline 
	\end{tabular}
	\caption{Mechanical properties of Nomex paper and UD fibre.\label{table_mat}}
	\end{center}
\end{table}

	The influence of a core's geometry on the transition frequency and modal density is the concern of this paper. Consequently, the mechanical properties and thicknesses of the skins, core and Nomex paper are identical in all the models. The sandwich panels developed below only differ in the geometry of their cores and the dimensions $d_x$, $d_y$ of their unit-cells. 
	
\begin{figure}[h!]
  \centering
  \includegraphics[width=0.24\linewidth]{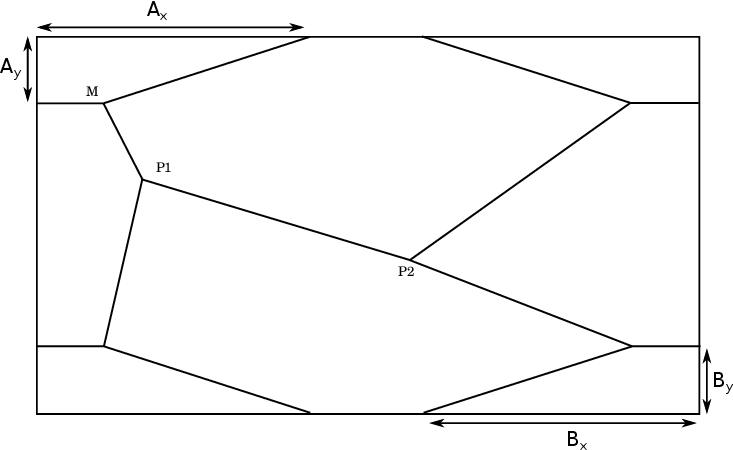}\hspace{0.02\linewidth}\includegraphics[width=0.26\linewidth]{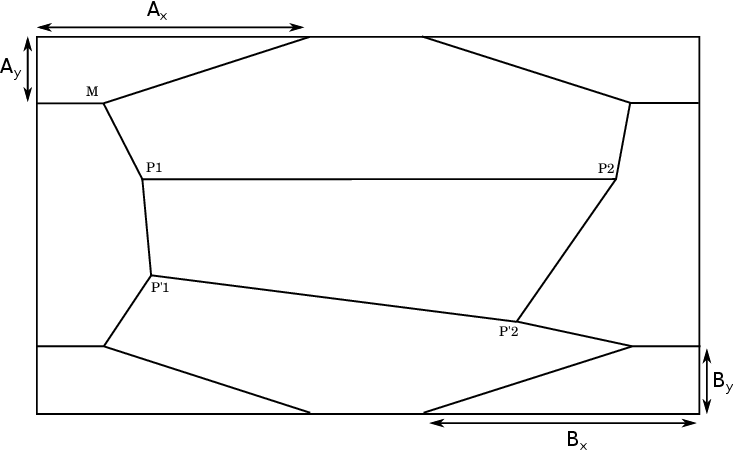}
  \caption{Top view of the 2D parametrized cores. Hexagonal (a) and octagonal (b) constructions.\label{parametric}}
\end{figure}

	The cores are created from the parametrized models shown in Figure \ref{parametric}. The rectangular boxes describe the periodic cell's dimensions, while the inner lines describe the top-view position of the Nomex paper in the core. A parametrized hexagonal core is proposed in Figure \ref{parametric}.a. The core construction is designed using the three points $M$, $P_1$, $P_2$ and the corner parameters $A_x$, $A_y$, $B_x$ and $B_y$. The other parameters are imposed by the compatibility requirement between the unit-cells. It can be seen that triangular, squares and pentagonal geometries can be obtained by setting coincident points $M = P_1$, $P_1 = P_2$ or $A_x = B_x$ and $A_y = B_y$.
	
	In Figure \ref{parametric}b, two additional points $P'_1$ and $P'_2$ are used for the construction of octagonal cores. Although this model can be easily generalised to create more complex shapes, it is reminded that the computational cost associated with optimisation procedures depends on the number of design parameters.
	
\begin{figure}[h!]
	\begin{center}
	\includegraphics[width=0.50\linewidth]{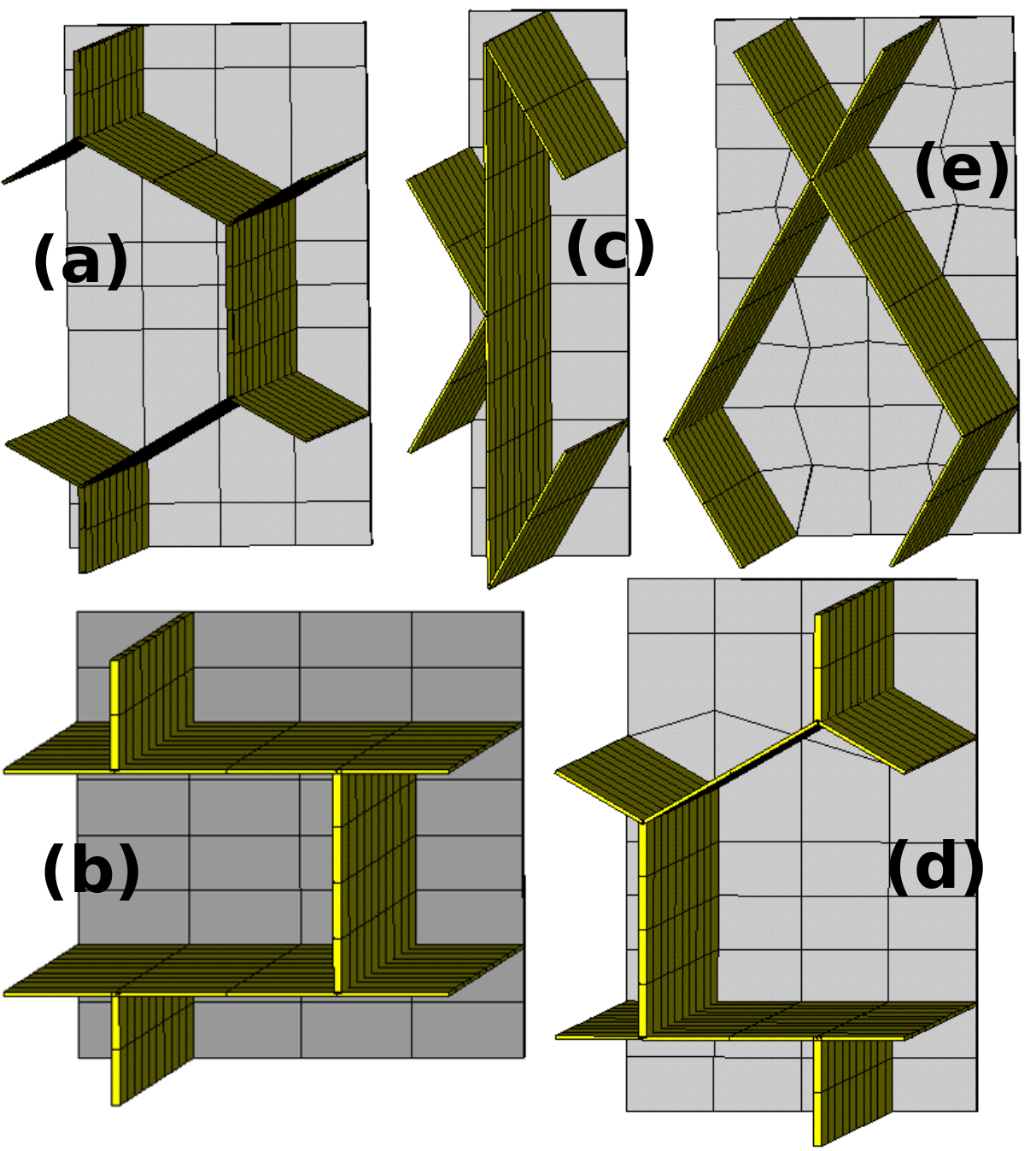} 
	\end{center}
	\caption{3D Finite element models of the considered periodic cells.\label{FEM}}
\end{figure}

\begin{figure}[h!]
  \centering
  \includegraphics[width=0.35\linewidth]{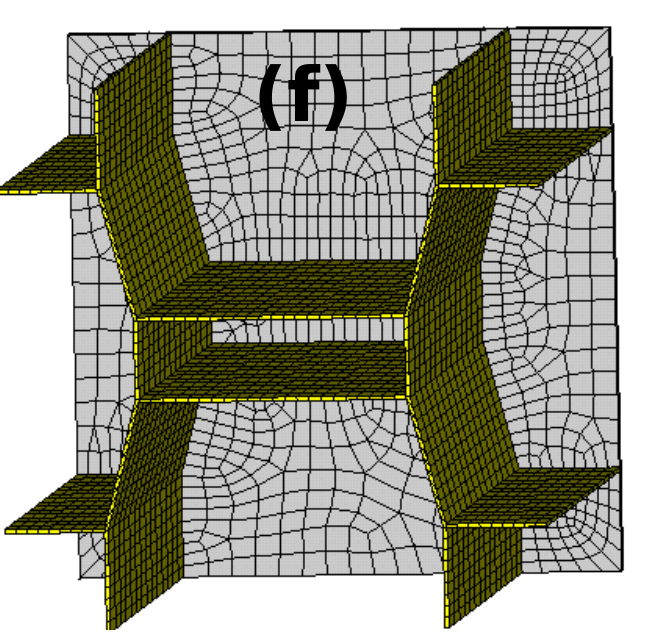}
  \caption{3D Finite element model of the octagonal construction.\label{FEM2}}
\end{figure}

	The finite element model of five unit-cells obtained using the first parametrized model is shown in Figure \ref{FEM}. These models involve linear block elements having 3 degrees of freedom per node. Case (a) is a classical hexagonal model (honeycomb). Case (b) is a rectangular, Case (c) is triangular, Case (d) is a pentagonal core and Case (e) is a parallelogram. A finely meshed octagonal core was also designed in Figure \ref{FEM2}. It should be noted that all the models are converged in the considered frequency range, but refined meshes such as in Figure \ref{FEM2} would be necessary in higher frequencies to describe the local dynamics at the cell's scale.
	
	Besides, the geometry of the core has a noticeable influence on the Nomex distribution, hence on the surface density of the sandwich panel. For this reason, the dimensions of the cellular core are chosen so that the overall bending stiffness-to-mass ratio remains constant in the sandwich. It means that the modal density is expected to be identical for all the considered models at low frequency. The size of the unit-cells, $d_x$ and $d_y$ are detailed in Table \ref{table_size}.

\begin{table}[!ht]
	\begin{center} 
	\begin{tabular}{|l|c|c|c|c|c|c|}
	\hline
	Case					&  (a)	&	(b)	& (c) 		& 	(d) 			& (e)  		& (f)\\
	\hline 
	$d_x$ (mm)		&  6.93	&		8	& 	4.29		& 		6.93		&	13.85 	& 20	 \\
	
	$d_y$ (mm)    	&  12	&		8	& 	14.48	&		10.53	&	8			& 20  \\
	\hline 
	\end{tabular}
	\caption{Size of the rectangular unit-cells in the six cases.\label{table_size}}
	\end{center}
\end{table}

	\subsection{Effects on the transition frequency}
	
\begin{figure}[ht]
	\begin{minipage}{1\columnwidth}
	\centering
	\includegraphics[width=0.42\linewidth]{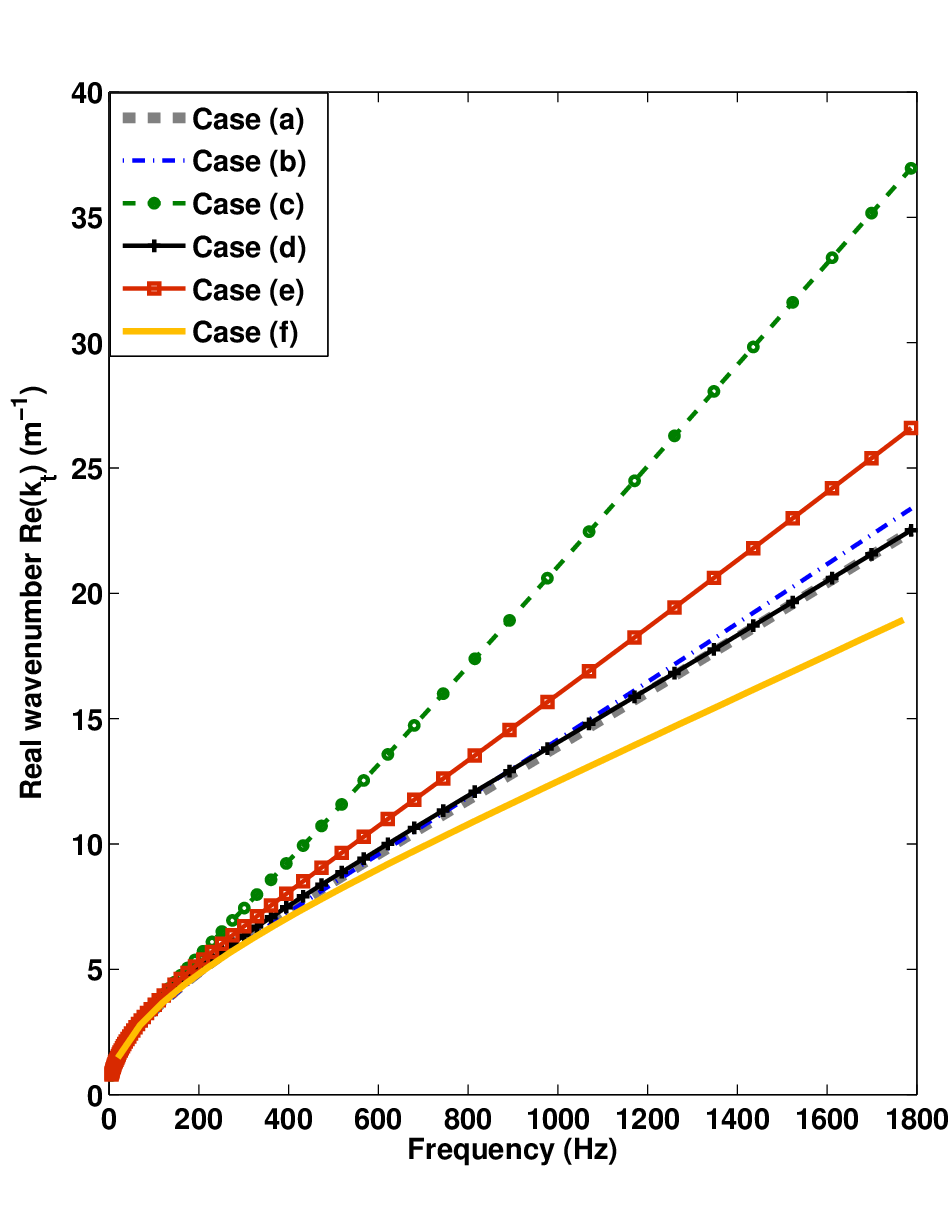}\hspace{0.08\linewidth}%
	\includegraphics[width=0.42\linewidth]{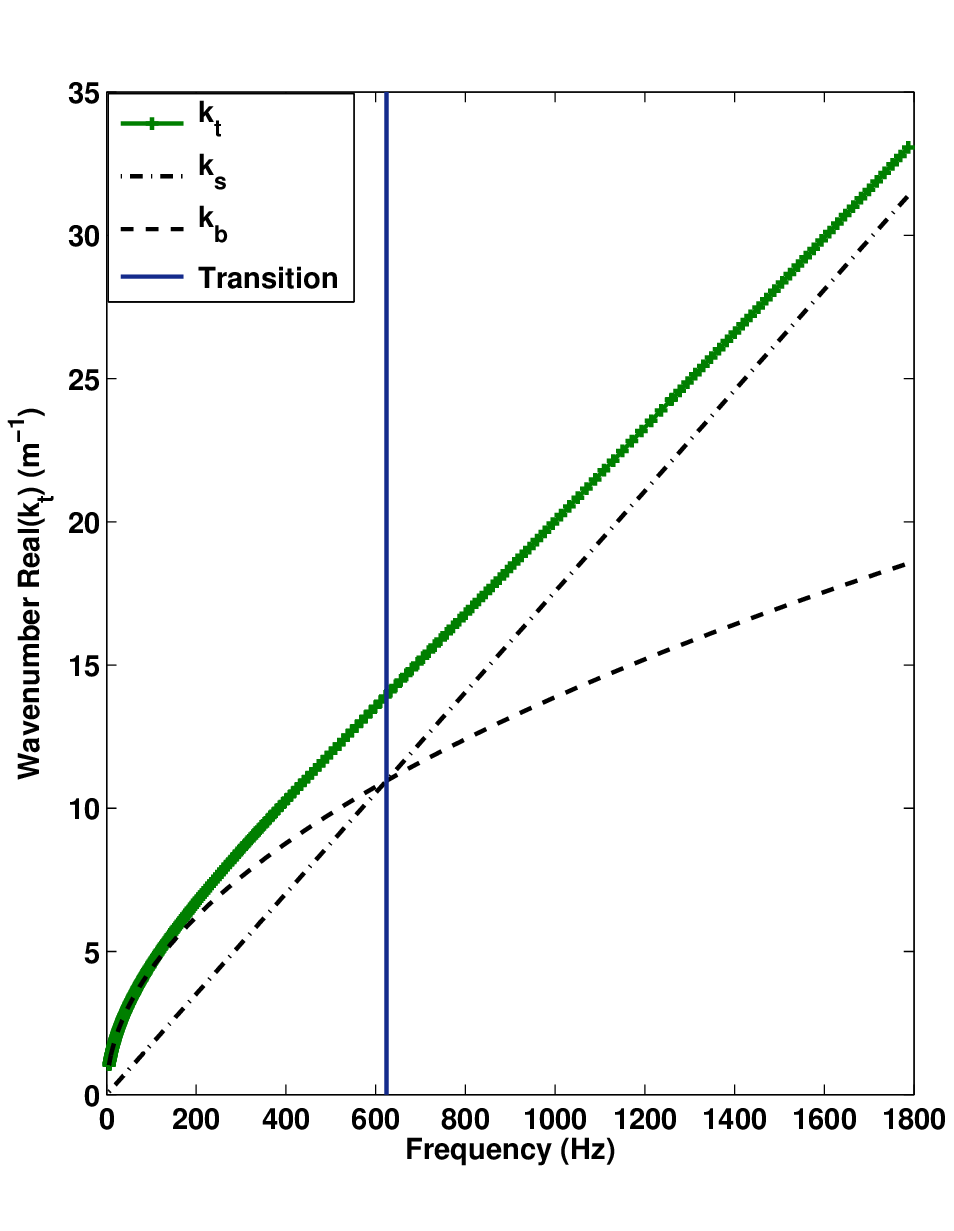}
	\end{minipage}
		\caption{Flexural wavenumbers for the six different core configurations (a) definition of the transition frequency: intersection between Case (b) asymptotic shear and bending waves (b).}
	\label{dispersion}
\end{figure}

	The wave dispersion characteristics of the periodic waveguides defined by unit-cells (a)-(f) are obtained using the method described in Section \ref{Sec2}. The flexural wavenumbers $k$ are compared in Figure \ref{dispersion}a. It is seen that the core geometry has a significant influence on the wavenumbers of the sandwich panel. More precisely, the flexural behaviours remain similar due to the constant bending rigidity and differ above the transition frequency. A 95\% difference can then be observed between the (c) and (f) wavenumbers at 1800\,Hz. 
	
	The transition frequency is located at the intersection between the asymptotic shear and bending curves. The method is shown in Figure \ref{dispersion}b for the core geometry (b), where the bending $k_b$ and shear $k_s$ wavenumbers are both derived from the WFEM solution. It should be noted that an accurate bending wavenumber requires a refined frequency sampling in the lower frequencies. For the considered models, it can be seen that a clear bending behaviour is observed below 200\,Hz, while the shear behaviour occurs above 600\,Hz except for case (f).
	
	For the honeycomb core (a), an exact analytical expression can be derived for the transition frequency. The dispersion relation in the sandwich can be written in terms of the mass per unit area $m_s$, the equivalent transverse shear stiffness $S$ and bending rigidity $D$ of the plate \cite{Guillaumie2015}:
	\begin{equation}
	k^2 = \dfrac{m_S}{2S} \left( \omega^2 + \omega \sqrt{\omega^2 + \dfrac{4 S^2}{m_S D}} \right)
	\label{eq_dispersion}
	\end{equation}
	
	Therefore, equalizing the asymptotic solutions defined Eqs.(\ref{eq_transitionK}) and (\ref{eq_transitionN}) yields:
	\begin{equation}
	\omega_{\textrm{T}} = \dfrac{S}{\sqrt{m_S D}}
	\end{equation}

	The transition frequency $\omega_{\textrm{T}}$ can therefore be expressed analytically in terms of the equivalent bending, shear and mass parameters. 
	
	For other cores' geometries, it should be noted that the transition frequency can be only approximated using Eqs.(\ref{eq_transitionK}) and (\ref{eq_transitionN}), since Gibson and Ashby homogenisation is no longer valid. The equivalent parameters $\tilde{m}_S$ ,$\tilde{D}$ and $\tilde{S}$ are obtained from Eq.(\ref{eq_dispersion}) and the numerical solution $k_t (\omega)$ evaluated using non-linear least-square regression. The error on the regression function is written:
	\begin{equation}
	\varepsilon_{\textrm{reg}}(\omega) = \left| \dfrac{k_t (\omega) - f_{\textrm{reg}}(\omega, \tilde{D}, \tilde{m}_S, \tilde{S})}{k_t (\omega)} \right|
	\end{equation}

\begin{figure}[ht]
  \centering
  \includegraphics[width=0.85\linewidth]{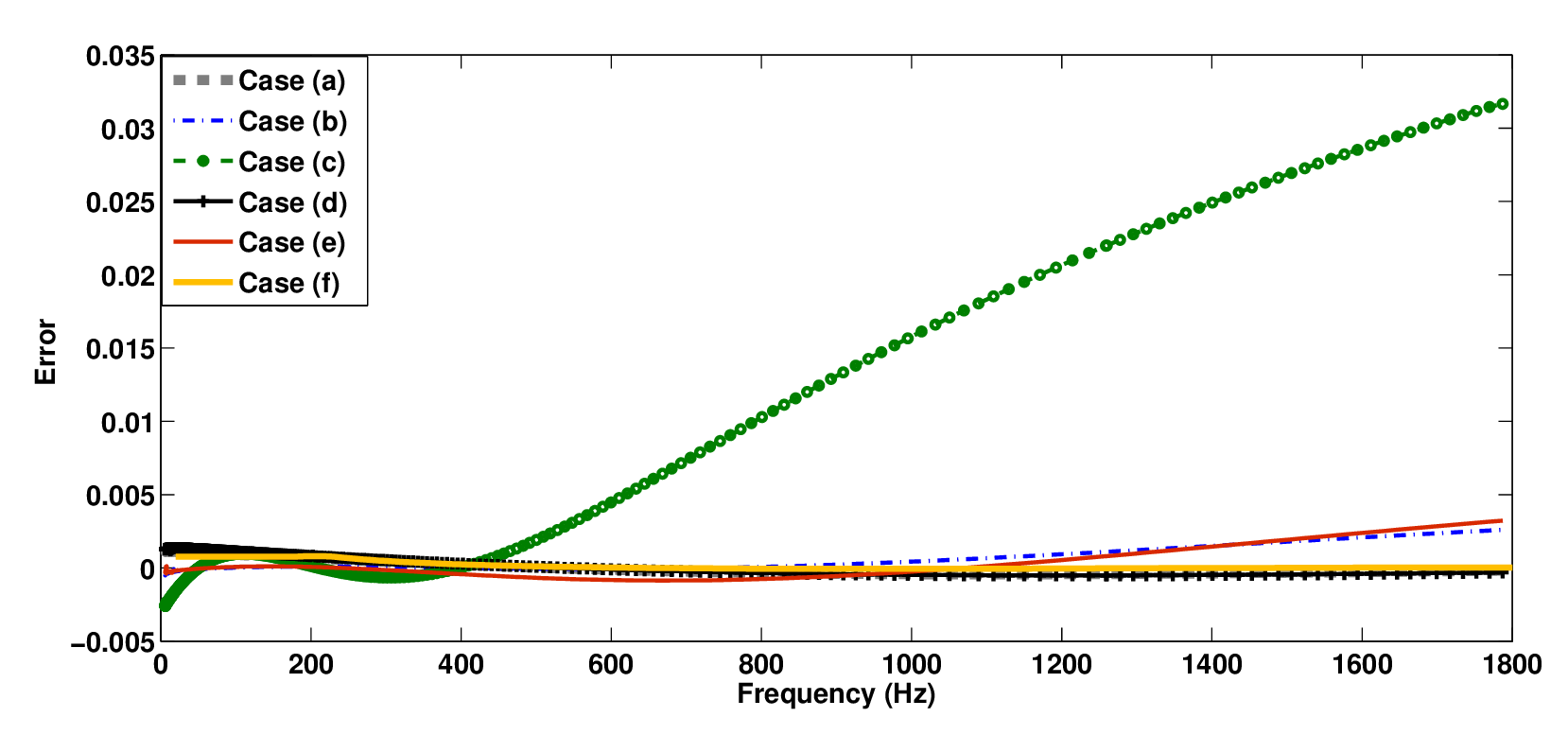}
  \caption{Difference between WFEM solutions and approximated wavenumbers.}
  \label{err_reg}
\end{figure}

	The values of $\varepsilon_{\textrm{reg}}$ are compared in Figure \ref{err_reg}. The regression appears in very good agreement with the WFEM solutions, except for Case (c) which exhibits up to 3\% error in the considered frequency range. On the other hand Case (d) and (f) fit very well with the honeycomb model described in Eq.(\ref{eq_dispersion}). It should be noted that the regression accuracy may depends on the chosen frequency range for $f_{\textrm{reg}}$.
			
	To summarize, the transition frequency can be determined from the asymptotic wavenumbers $k_s$ and $k_b$ derived from the WFEM solutions. In order to reduce the sensitivity to sampling for the determination of $k_b(\omega \rightarrow 0)$, the equivalent shear, bending and mass parameters can be obtained through a regression model. An analytical expression can therefore be derived for $\omega_{\textrm{T}}$ with a good accuracy, except for case (c).

\subsection{Influence on vibro-acoustic parameters}

\begin{figure}[ht]
	\begin{center} 
	\includegraphics[width=0.45\linewidth]{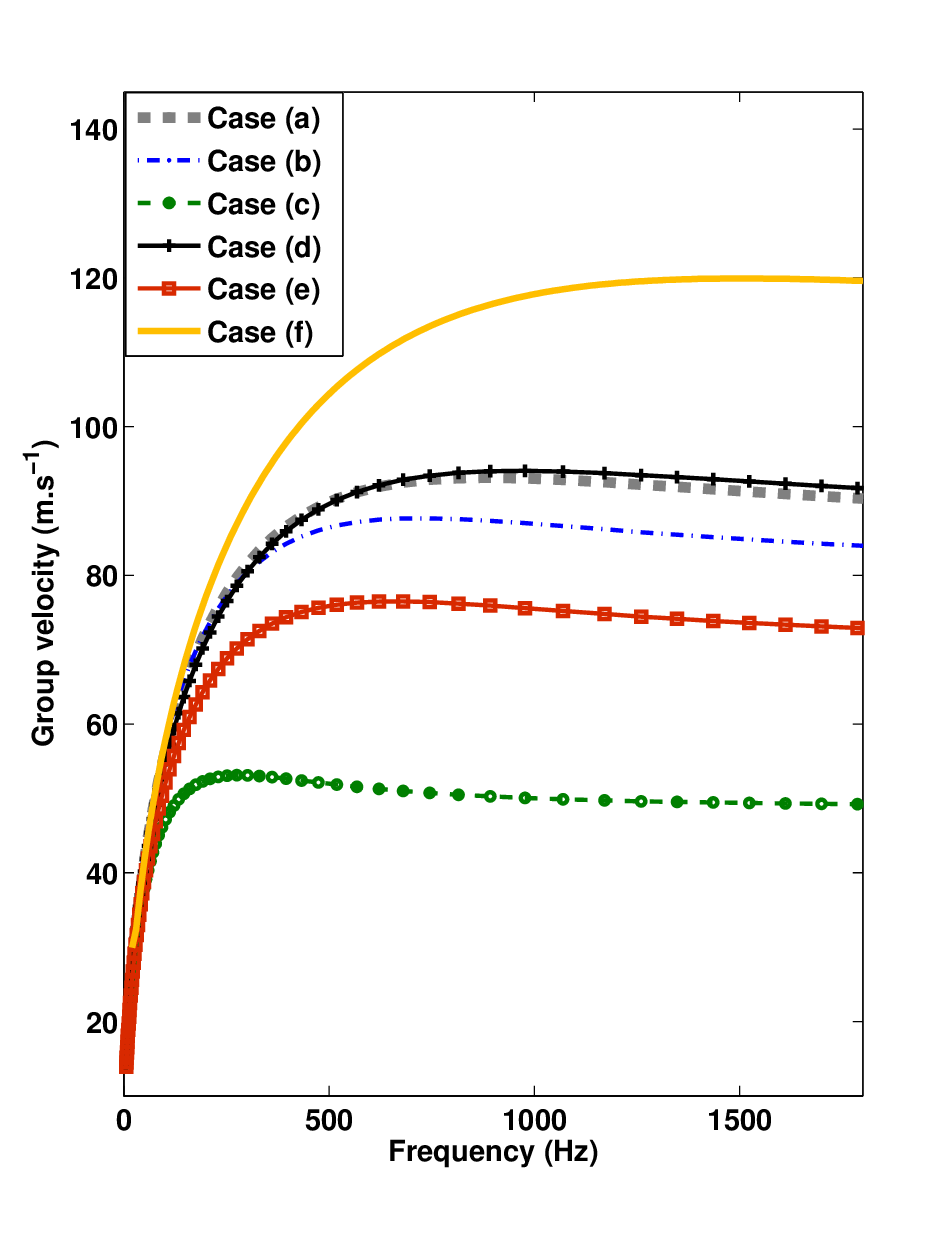} 
	\caption{Averaged energy velocity in the sandwich waveguide.\label{cg_equiv}}
	\end{center}
\end{figure}

The transition can be observed on the averaged group velocity $c_g$, as the passage from a square to linear function. The velocities are compared in Figure \ref{cg_equiv}. It can be seen that the cores have a significant influence on the energy velocity in the panel, and the waves are still dispersive above the transition frequency. Noteworthy, the maximal group velocity is obtained around the transition, meaning that reduced wave attenuations are expected in this frequency range.

	The transition frequencies are detailed in Table \ref{table_transition} and compared with the honeycomb model (a). Case (c), associated with triangular construction has the lowest shear rigidity. The rectangular and parallelogram cores, cases (b) and (e) are associated with reduced transition frequencies compared to the hexagonal (a) core. The pentagonal core (d), produces a slight increase (9.11\,\%) of the transition frequency compared with honeycomb. On the other hand, the octagonal core (f) provides a considerable increase (70\,\%) of the transition frequency. Therefore, this geometry can postpone the transition frequency while maintaining the stiffness-to-mass ratio.

	\begin{table}[!ht]
	\begin{center} 
	\begin{tabular}{|l|c|c|c|c|c|c|}
	\hline
	Case					&  (a)	&	(b)	& (c) 		& 	(d) 			& (e)  & (f)\\
	\hline 
	Transition frequency (Hz) & 765.79 & 624.03 &  244.64 &  835.54 & 570.29 & 1302.5 \\
	\hline 
	Variation with honeycomb & $\diagup$ & - 18.5 \% & - 68 \% & \textbf{+ 9.11 \%} & - 25.5 \% & \textbf{+ 70.05 \%} \\
	\hline
	\end{tabular}
	\caption{Transition frequency in the panel for different cores geometries.\label{table_transition}}
	\end{center}
	\end{table}
	
\begin{figure}[ht]
  \centering
  \includegraphics[width=0.80\linewidth]{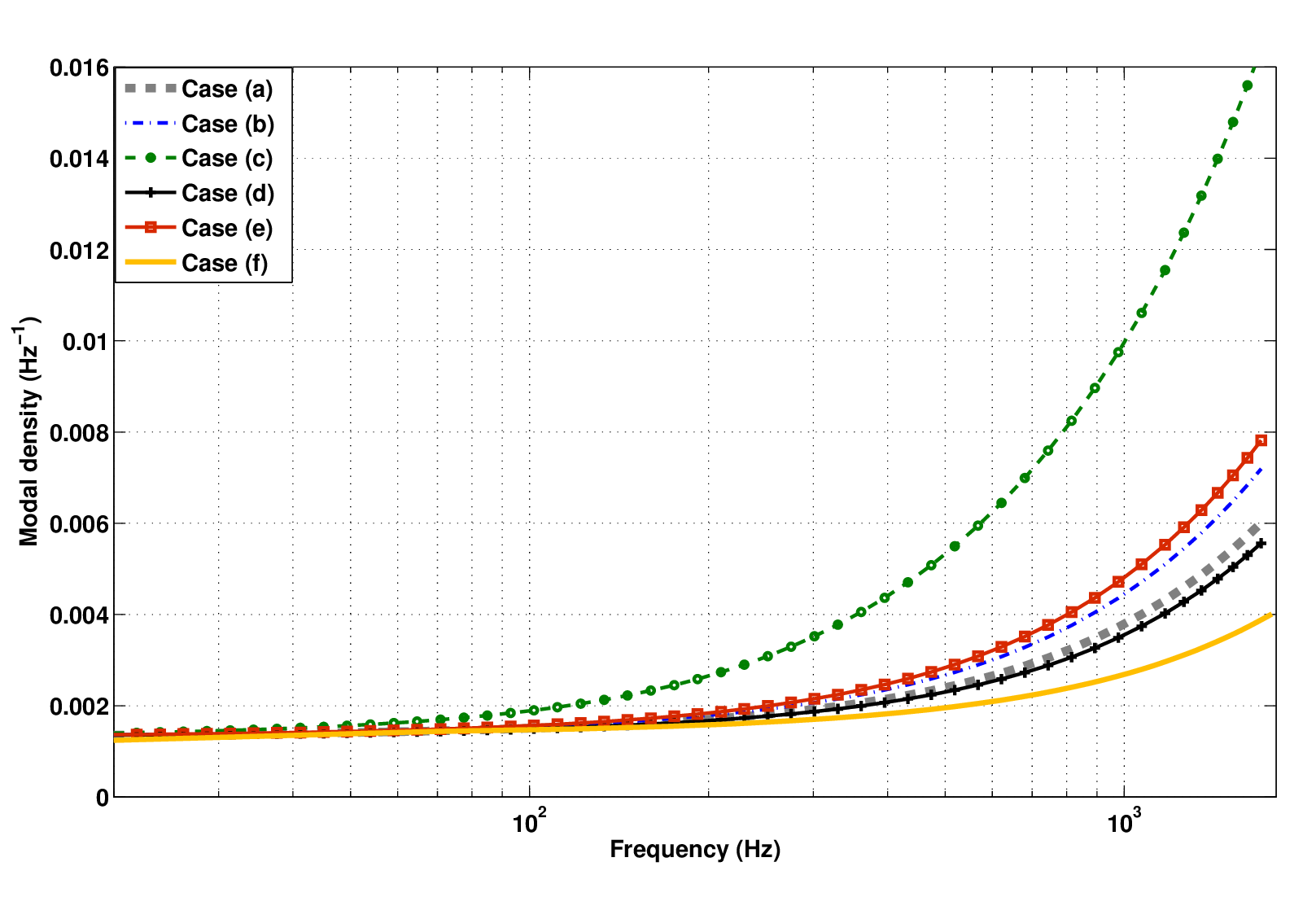}
  \caption{Modal density in the sandwich panels with different cores.\label{modal_dens}}
\end{figure}
	
	 The modal density in the sandwich panel is calculated from the propagating wavenumbers and group velocities as mentioned above. The WFEM solutions are compared in Figure \ref{modal_dens} for a surface S=1\,m$^2$, although it is reminded that the transition frequency does not depend on the panel's surface. The scaling procedure to preserve the mass-to-stiffness ratios is validated since the modal densities $n(\omega \rightarrow 0)$ are the same for the six models. It can be seen that $n(\omega)$ critically increases above $\omega_{\textrm{T}}$ for each model. It should be noted that the influence of the core's geometry on the overall flexural behaviour will increase if the ratio between the core $h_T$ and the skins $h_s$ increases. 

	\section{Concluding remarks}
	
This paper is concerned with the design of sandwich panels with improved vibro-acoustic parameters. Although honeycomb constructions provide excellent compressional strength to density ratio, other types of periodic cores can be used to increase the transition frequency and reduce the modal density in a given frequency range. 

In this work, the WFEM was employed to compare various types of core constructions for a given sandwich panel. This method does not suffer the limitations associated with core homogenisation and provides accurate wave dispersion characteristics with a reasonable computational cost. Additionally, the method can be further used to compute bandgaps and local resonances of the periodic cells which have a significant influence on the acoustic parameters in higher frequency (see Song et al.\cite{Song2015}). It was shown that a 70\,\% increase of the transition frequency can be obtained by using an octagonal core geometry, while maintaining the same bending stiffness-to-mass ratio, as well as the skin and core materials. As a consequence of the increased equivalent shear modulus, the octagonal core also produces significantly higher group velocity and reduced modal density.

This transition frequency can be used as a relevant output for optimisation procedures oriented toward improving vibro-acoustic parameters. It should be noted that the proposed application is restrained to 2D geometric parameters (the core is unchanged along its thickness), since a reduced number of design parameters is required for sensitivity of optimisation procedures. An interesting development would be to understand how 3D cores could be designed in order to significantly postpone the bending to shear transition to higher frequencies.

\begingroup\renewcommand{\doi}[1]{doi:#1}

\endgroup


\begin{thebibliography}{21}
\providecommand{\natexlab}[1]{#1}
\providecommand{\url}[1]{\texttt{#1}}
\providecommand{\urlprefix}{URL }
\expandafter\ifx\csname urlstyle\endcsname\relax
  \providecommand{\doi}[1]{doi:\discretionary{}{}{}#1}\else
  \providecommand{\doi}{doi:\discretionary{}{}{}\begingroup
  \urlstyle{rm}\Url}\fi
\providecommand{\eprint}[2][]{\url{#2}}
\providecommand{\BIBand}{and}
\providecommand{\bibinfo}[2]{#2}
\ifx\xfnm\undefined \def\xfnm[#1]{\unskip,\space#1}\fi
\bibitem[{Reddy(1997)}]{Reddy1997}
Reddy JN. Mechanics of laminated composite plates--Theory and analysis. Boca Raton, FL: CRC Press; 1997.
\bibitem[{Kurtze and Watters(1959)}]{Kurtze1959}
Kurtze G, Watters BG. New wall design for high transmission loss or high damping. J Acoust Soc Am 1959;31(6):739--48.
\bibitem[{Wilkinson(1968)}]{Wilkinson1968}
Wilkinson J. Modal densities of certain shallow structural elements. J Acoust Soc Am 1968;43(2):245--51.
\bibitem[{Erickson(1969)}]{Erickson1969}
Erickson LL. Modal densities of sandwich panels: theory and experiment. Shock Vib Bull 1969;39(3):1--16.
\bibitem[{Renji et~al.(1996)Renji, Nair and Narayanan}]{Renji1996}
Renji K, Nair P, Narayanan S. Modal density of composite honeycomb sandwich panels. J Sound Vibr 1996;195(5):687--99.
\bibitem[{Ghinet and Atalla(2011)}]{Ghinet2011}
Ghinet S, Atalla N. Modeling thick composite laminate and sandwich structures with linear viscoelastic damping. Comput Struct 2011;89(15):1547--61.
\bibitem[{Guillaumie(2015)}]{Guillaumie2015}
Guillaumie L. Vibroacoustic flexural properties of symmetric honeycomb sandwich panels with composite faces. J Sound Vibr 2015;343:71--103.
\bibitem[{Florence and Renji(2015)}]{Florence2015}
Florence SJK, Renji K. Modal density of thin composite cylindrical shells. J Sound Vibr 2015. \url{http://dx.doi.org/10.1016/j.jsv.2015.11.030i}.
\bibitem[{Gibson and Ashby(1997)}]{Gibson1997}
Gibson LJ, Ashby MF. Cellular solids: structure and properties. Cambridge University Press; 1997.
\bibitem[{Mace et~al.(2005)Mace, Duhamel, Brennan and Hinke}]{MaceJASA2005}
Mace BR, Duhamel D, Brennan MJ, Hinke L. Finite element prediction of wave motion in structural waveguides. J Acous Soc Am 2005;117:2835--43.
\bibitem[{Mencik and Ichchou(2005)}]{MencikEJM2005}
Mencik JM, Ichchou MN. Multi-mode propagation and diffusion in structures through finite elements. Eur J Mech A-Solid 2005;24(5):877--98.
\bibitem[{Alkhader et~al.(2015)Alkhader, Iyer, Shi and
  Venkatesh}]{Alkhader2015}
Alkhader M, Iyer S, Shi W, Venkatesh T. Low frequency acoustic characteristics of periodic honeycomb cellular cores: the effect of relative density and strain fields. Compos Struct 2015;133:77--84.
\bibitem[{Catapano and Montemurro(2014)}]{Catapano2014}
Catapano A, Montemurro M. A multi-scale approach for the optimum design of sandwich plates with honeycomb core. part ii: the optimisation strategy. Compos Struct 2014;118:677--90.
\bibitem[{Casadei et~al.(2016)Casadei, Rimoli and Ruzzene}]{Ruzzene2016}
Casadei F, Rimoli J, Ruzzene M. Multiscale finite element analysis of wave propagation in periodic solids. Finite Elem Anal Des 2016;108:81--95.
\bibitem[{Zhou et~al.(2015)Zhou, Lain{\'e}, Ichchou and Zine}]{Zhou2015CS}
Zhou C, Lainé J, Ichchou M, Zine A. Multi-scale modelling for two-dimensional periodic structures using a combined mode/wave based approach. Comput Struct 2015;154:145--62.
\bibitem[{Droz et~al.(2014)Droz, Lain{\'e}, Ichchou and
  Inqui{\'e}t{\'e}}]{DrozCS2014}
Droz C, Lainé JP, Ichchou M, Inquiété G. A reduced formulation for the free wave propagation analysis in composite structures. Compos Struct 2014;113:134--44.
\bibitem[{Bloch(1929)}]{Bloch1929}
Bloch F. Über die quantenmechanik der elektronen in kristallgittern. Zeitschrift für physik 1929;52(7--8):555--600.
\bibitem[{Droz et~al.(2016)Droz, Zhou, Ichchou and Lain{\'e}}]{Droz2015}
Droz C, Zhou C, Ichchou M, Lainé JP. A hybrid wave-mode formulation for the vibro-acoustic analysis of 2d periodic structures. J Sound Vib 2016;363:285--302.
\bibitem[{Lesueur et~al.(1988)Lesueur, Delcambre and Heckl}]{Lesueur1988}
Lesueur C, Delcambre J, Heckl M. Rayonnement acoustique des structures: vibroacoustique, interactions fluide-structure. Eyrolles 1988.
\bibitem[{Langley(1994)}]{Langley1994}
Langley R. The modal density and mode count of thin cylinders and curved panels. J Sound Vib 1994;169(1):43--53.
\bibitem[{Song et~al.(2015)Song, Feng, Wen, Yu and Wen}]{Song2015}
Song Y, Feng L, Wen J, Yu D, Wen X. Reduction of the sound transmission of a periodic sandwich plate using the stop band concept. Compos Struct 2015;128:428--36.
\end{thebibliography}
\end{document}